\documentstyle[aps,prd]{revtex}

\begin{document}

\def \ts {\textstyle}
\def \rd {\displaystyle{\cdot}}
\def \D {\mbox{D}}
\def \ep {\varepsilon}
\def \la {\langle}
\def \ra {\rangle}
\def \c {\mbox{curl}\,}
\def \p {\partial}
\def \cs {c_{\rm s}^2}
\input epsf
\def\plot#1{\centering \leavevmode
\epsfxsize= 1.0\columnwidth \epsfbox{#1}}
\newcommand{\be}{\begin{equation}}
\newcommand{\ee}{\end{equation}}
\newcommand{\ben}{\begin{eqnarray}}
\newcommand{\een}{\end{eqnarray}}
\newcommand{\n}{\label}
\newcommand{\no}{\noindent}
\newcommand{\lsim}{\mbox{\raisebox{-.3em}{$\stackrel{<}{\sim}$}}}

\title{Quintessence dissipative superattractor cosmology}

\author{
Luis P. Chimento,  Alejandro S. Jakubi and Norberto A. Zuccal\'a}

\address{Departamento de F\'{\i}sica, Universidad de
Buenos Aires, 1428~Buenos Aires, Argentina}

\maketitle
\date{\empty}
\begin{abstract}

We investigate the simplest quintessence dissipative dark matter attractor
cosmology characterized by a constant quintessence baryotropic index and a
power--law expansion. We show a class of accelerated coincidence--solving
attractor solutions converging to this asymptotic behavior. Despite its
simplicity, such a ``superattractor'' regime provides a model of the recent
universe that also exhibits an excellent fit to supernovae luminosity
observations and no age conflict. Our best fit gives $\alpha=1.71\pm 0.29$ for
the power-law exponent. We calculate for this regime the evolution of density
and entropy perturbations.

\end{abstract}
\  \\
\  \\
\  \\
\noindent
PACS number(s): 98.80.Hw, 04.20Jb

\newpage

\section{Introduction}

A new picture of the universe is emerging from observations of large scale
structure, searches for Type Ia supernovae, and measurements of the cosmic
microwave background anisotropy. They all suggest that the universe is
undergoing cosmic acceleration and is dominated by a smoothly distributed dark
energy component with large negative pressure \cite
{Bahcall99,Turner00,Primack00}. Moreover, recent data coming from BOOMERANG
and MAXIMA projects seem to indicate that the density of the universe is near
the critical density \cite{CMB}. The most frequently proposed candidates for
this dark energy are a cosmological constant, or vacuum energy density \cite
{lambda}, and quintessence \cite{quint}, a scalar field with negative
pressure. On the other hand, gravitationally clustered matter is usually
assumed to be dominated by cold, collisionless, nonbaryonic dark matter (CDM).

In addition to the old cosmological constant problem, related to the smallness
of the observational upper bound on the vacuum energy density compared to
particle physics scales \cite{Weinberg89}, a new challenge to the model with
cosmological constant and CDM ($\Lambda$CDM) is the ``cosmic coincidence''
problem: why is it that the vacuum density dominates the universe only
recently \cite{Stein97}. A dynamical self--interacting scalar field may
explain why the dark energy is small. However, a minimally coupled scalar
field combined with perfect fluid matter, such as in ``tracking'' QCDM models
\cite{track}, cannot explain the observed acceleration and solve the
coincidence problem \cite{cjp,Weinberg00}. On the other hand it has been shown
in Ref. \cite{cjp} that both acceleration and coincidence can be
satisfactorily explained by a combination of quintessence and dissipative dark
matter (QDDM). For these models it was shown that late--time attractor
solutions exist with very interesting properties: an accelerated expansion,
spatially flatness, and a fixed ratio of quintessence to dark matter energy
density. Recently, it has been shown that such accelerated
coincidence--solving attractor solutions also exist in spatially flat models
with an exponential self-interaction quintessence potential and
phenomenologically chosen couplings between the quintessence field and
baryotropic dark matter \cite{Coley00,Amendola}.

Consideration of dissipative effects in dark matter also arises from
increasing evidence that numerical simulations of dark matter halos on
sub-galactic scales based on conventional CDM models lead to conflicts with
observations. One of the major problems shown by these simulations is that
galactic halos are too centrally concentrated \cite{halos}. Confirmation of
this problem would imply that structure formation is somehow suppressed on
small scales. Several scenarios addressing this issue have been considered
assuming some kind of interaction for dark matter particles: nonthermally
produced and weakly interacting \cite{Lin00}, self--interacting \cite{self},
repulsive \cite{Goodman00}, annihilating \cite{ann}, and decaying \cite{Cen}.
It is quite reasonable to expect that dark matter is out of thermodynamical
equilibrium and these same interactions are at the origin of the cosmological
dissipative pressure. A simple estimation shows that a cross section of the
order of magnitude proposed in these halo formation scenarios, corresponding
to a mean free path in the range $1$ kpc to $1$ Mpc, yields
at cosmological densities a mean free  path a bit lower than the Hubble
distance. Hence a description for interacting dark matter as a dissipative
fluid is valid at cosmological scales \cite{Pavon93}.

Dark matter may also have several components, from very heavy weakly
interacting massive particles to a lightweight neutrino \cite{Gross}. Even if
this neutrino-like component does not contribute significantly to the density
budget of the universe \cite{Primack00}, it may have nevertheless a relevant
dynamical role as distinct components of dark matter, cooling at different
rates, give rise to bulk viscosity \cite{Zimdahl96}.

Recently the relationship between dark matter clustering, the nature of dark
matter and the origin of ultrahigh energy cosmic rays (UHECRs) has been
explored \cite{Blasi}. Heavy particles ($m_X\sim 10^{12}-10^{14}$ GeV) can be
produced in the early universe in different ways \cite{kolb,bere,kuzmin,kt}
and their lifetime can be finite though longer than the age of the universe.
In these circumstances, super-heavy particles can represent an appreciable
fraction of dark matter and the decay of these particles results in the
production of UHECRs, as widely discussed in the literature
\cite{bere,sarkar,bbv,blasi}. In particular, if the relics cluster in galactic
halos, as is expected, they can explain the cosmic ray observations above
$\sim 5\times 10^{19}$ eV. The effect of decaying dark matter on the equation
of state has been studied numerically in Ref. \cite{Ziae}, and the equivalence
between particle production and dissipative bulk viscosity has been
investigated in Ref. \cite{Zimdahl00}.

All this shows that many different scenarios may occur where significant
dissipative processes develop in dark matter, in particular when it behaves as
a viscous fluid. So a variety of accelerated coincidence--solving attractors
are possible. In this paper we will show that for a wide class of QDDM models
attractor solutions are themselves attracted towards a common asymptotic
behavior, the ``superattractor''. The superattractor scenario is described in
Sec. II and observational constraints on this scenario are investigated in
Sec. III. These include the luminosity distance--redshift relation for type Ia
supernovae and the age of the universe. We apply the covariant
gauge--invariant formalism to calculate the evolution of density and entropy
scalar perturbations in the QDDM regime and solve them for the superattractor
regime in Sec. IV. Finally the main conclusions are discussed in Sec. V. Units
in which $c = 8\pi G = k_{B} = 1$ are used throughout.

\section{QDDM superattractor scenario}

In Ref. \cite{cjp} it was shown that the
Friedmann--Lamaitre--Robertson--Walker (FLRW) universe filled with perfect
normal matter plus quintessence fluid, corresponding to some minimally coupled
scalar field governed by the Klein--Gordon equation, cannot at the same time
drive an accelerated expansion and solve the coincidence problem. To solve it,
some additional bulk dissipative pressure $\pi$ in the stress--energy tensor
of dark matter was considered. Any dissipation in FLRW universes has to be
scalar in nature, and in principle it may be modeled as a bulk viscosity
effect within a nonequilibrium thermodynamic theory such as the
Israel--Stewart theory \cite{d1,m2}. In a certain regime, that formulation
can be approximated by the more manageable truncated transport equation

\begin{equation}
\pi + \tau\dot{\pi}  =  - 3\zeta H \, ,
\label{dpi2}
\end{equation}

\no where $H \equiv \dot{a}/a$ denotes the Hubble factor, $\zeta $ stands for
the phenomenological coefficient of bulk viscosity, and $\tau$ is the
relaxation time associated with the dissipative pressure \cite{mdv,jpa,HL}. As
usual an overdot means derivative with respect to cosmic time.

The overall stress--energy tensor of the QDDM model reads

\begin{equation}
T_{ij} = (\rho_m+\rho_{\phi} + p_m+p_{\phi} +\pi) u_{i} u_{j} +
(  p_m+p_{\phi}+\pi) g_{ij},
\label{2}
\end{equation}

\noindent where $\rho = \rho_{m} + \rho _{\phi}$  and $p = p_{m} + p_{\phi}$.
Here $\rho_m$ and $p_m$ are the energy density and pressure of the matter
whose equation of state is $p_{m} = (\gamma_{m} -1) \rho_{m}$ with baryotropic
index in the interval $ 1 \leq \gamma_{m} \le 2$. Likewise $\rho_{\phi}$ and
$p_{\phi}$, the energy density and pressure of the minimally coupled
self--interacting quintessence field $\phi$, are related by the equation of
state, $p_{\phi} = (\gamma_{\phi} - 1) \rho_{\phi}$, with baryotropic index

\begin{equation} \label{gammaphi}
\gamma_\phi=\frac{\dot\phi^2}{\dot\phi^2/2+V(\phi)},
\end{equation}

\noindent where for non-negative potentials $V(\phi)$ one has $0 \leq
\gamma_{\phi} \leq 2$. The scalar field can be properly interpreted as
quintessence provided $\gamma_{\phi} < 1$ --see e.g. \cite{quint}. In general
$\gamma_{\phi}$ varies as the universe expands, and the same is true for
$\gamma_{m}$ since the massive and massless components of the matter fluid
redshift at different rates. 

The Friedmann equation and the energy conservation of the normal
matter fluid and quintessence (Klein--Gordon equation) are

\be
\label{feq}
H^{2}  + \frac{k}{a^{2}} =  \frac{1}{3}( \rho_{m} + \rho_{\phi}) 
\qquad (k = 1, 0, -1),
\ee

\be
\dot{\rho_m}  +  3 H\left(\gamma_m+\frac{\pi}{\rho_m}\right)\rho_m  = 0 ,
\label{drho}
\ee

\be
\dot{\rho_{\phi}}+ 3H\gamma_{\phi}\rho_{\phi}= 0 ,
\label{KG}
\ee

\noindent where the prime
means derivative with respect to $\phi$. Introducing $\Omega_m \equiv \rho_{m}
/\rho_{c}$, $\Omega_{\phi} ,\equiv \rho_{\phi} /\rho_{c}$, with $\rho_{c}
\equiv 3 H^{2}$ the critical density and $\Omega_k \equiv-k/(aH)^2$ plus the
definition $\Omega \equiv \Omega_{m} + \Omega_{\phi}$, the set of equations
(\ref{feq})--(\ref{KG}) can be recast as (cf. \cite{Ellis})

\be
\label{constr}
\Omega_{m} + \Omega_{\phi}+\Omega_k=1,
\ee

\be
\label{dOmegam}
\dot{\Omega}_{m}+3H\left(\gamma_m+\frac{\pi}{\rho_{m}}+
\frac{2\dot H}{3H^2}\right)\Omega_m=0 \, ,
\ee

\be
\label{dOmegaphi}
\dot{\Omega}_{\phi}+3H\left(\gamma_{\phi}+\frac{2\dot
H}{3H^2}\right)\Omega_{\phi}=0 \, ,
\ee

\noindent
where $\gamma$ is the effective baryotropic index given by

\begin{equation} \label{gammaOmega}
\gamma\Omega=\gamma_m\Omega_m+\gamma_\phi\Omega_\phi.
\end{equation}

Equations (\ref{constr})--(\ref{dOmegaphi}) have fixed point
solutions $\Omega=1$, $\Omega_m=\Omega_{ma}$ and $\Omega_\phi=\Omega_{\phi
a}$, respectively, when the partial baryotropic indices and the dissipative
pressure are related by

\begin{equation} \label{gammapi}
\gamma_{m}+\frac{\pi}{\rho_{m}}=\gamma_{\phi} = -\frac{2\dot{H}}{3H^2}.
\end{equation}

\noindent Asymptotical stability of $\Omega=1$ occurs whenever
$\gamma+\pi/\rho<2/3$. This condition, together with Eq. (\ref{gammapi}),
leads to the additional constraints $\pi<\left(2/3-\gamma_m\right)\rho_{m}$,
which is negative for ordinary matter fluids, and $\gamma_{\phi}<2/3$.
Additionally the stability of $\Omega_m=\Omega_{ma}$ and
$\Omega_\phi=\Omega_{\phi a}$, and hence of the ratio $\Omega_m/\Omega_\phi$,
has been established in \cite{cjp}. These solutions provide a natural
explanation to several  features observed in our universe: an accelerated
expansion, spatial flatness, and a ratio of dark energy to matter density of
order unity. We denote by a subindex $a$ the asymptotic limit of magnitudes in
the attractor regime, while the subindex $0$ will denote as usual their
current values.

Combining Eq. (\ref{gammapi}) with Eq. (\ref{dpi2}) we obtain
the equation of motion of the attractor
solutions of the system (\ref{feq}), (\ref{drho}), (\ref{KG}), (\ref{dpi2})
satisfying flatness, acceleration and coincidence:

\begin{equation} \label{dHa}
\nu^{-1}\left(\frac{\ddot H}{H}+3\gamma_m \dot H\right)+
\dot H+\frac{3\gamma_m}{2}H^2
-\frac{3\zeta}{2\Omega_{ma}}H=0 \,.
\end{equation}

\no Here $\nu=\left ( \tau H\right )^{-1}$ is the number of relaxation times
in a Hubble time -- for a quasistatic expansion $\nu$ is proportional to the
number of particle interactions in a Hubble time. Perfect fluid behavior
occurs in the limit $\nu\to\infty$, and a consistent hydrodynamical
description of the fluids requires $\nu>1$.  Rewriting (\ref{dHa}) in terms of
the field baryotropic  index $\gamma_{\phi}$, we get

\be
\n{gpa}
\gamma'_{\phi}=3\gamma_{\phi}^2-(\nu+3\gamma_m)\gamma_{\phi}+
\nu\left(\gamma_m-\frac{\zeta}{\Omega_{ma}H}\right)\,,
\ee

\no where a prime indicates derivative with respect to $\eta=\ln a$.

When the
phenomenological coefficient of bulk viscosity satisfies $\zeta=\zeta_s$,
where

\be
\n{vis}
\zeta_s=\Omega_{ma}\left(\gamma_m-\gamma_{\phi s}\right)\left[1-3\gamma_{\phi s}
\,\nu^{-1}\right]H
\equiv \kappa H \,,
\ee

\noindent Eq. (\ref{gpa}) admits the constant solution
$\gamma_\phi=\gamma_{\phi s}$. It gives an accelerated expansion in the late
time regime when $\gamma_{\phi s}<2/3$. As $\zeta>0$ and $\gamma_m\ge 1$, the
hydrodynamical parameter $\nu$ is restricted to $\nu>3\gamma_{\phi s}$. The
case of constant $\kappa$ in the interval $0 < \kappa <1$ arises, for
instance, in a radiating fluid, and the nearly linear regime, with slowly
varying $\nu$ and $\gamma_m$, was already investigated in the quasiperfect
limit, corresponding to $\nu^{-1}\to 0$ \cite{cjp}.

To analyze the stability of the solution $\gamma_\phi=\gamma_{\phi s}$ we
insert Eq. (\ref{vis}) into (\ref{gpa})

\be
\n{gpaa}
\gamma'_{\phi}=3\left(\gamma_{\phi}^2-\gamma_{\phi s}^2\right)
-(\nu+3\gamma_m)\left(\gamma_{\phi}-\gamma_{\phi s}\right) \,.
\ee

\no As $\gamma_{\phi s}<2/3$, $\nu>\mbox{\rm max}(3\gamma_{\phi s},1)$, and
$\gamma_m\ge 1$, Eq. (\ref{gpaa}) shows that
$\partial\gamma_\phi'/\partial\gamma_\phi<0$ in a neighborhood of
$\gamma_{\phi s}$. Hence this constant solution is asymptotically stable,
showing that all solutions of Eq. (\ref{dHa}), that is, all the accelerated
coincidence--solving attractors of the system (\ref{feq}), (\ref{drho}),
(\ref{KG}), (\ref{dpi2}), are themselves attracted towards the constant
solution $\gamma_\phi=\gamma_{\phi s}$ provided that they satisfy $\zeta\sim
\zeta_s$ when $t\to\infty$. As the same occurs with all solutions whose
initial conditions lay within the domain of attraction of each of these
attractor solutions, we will refer to the constant solution as the
``superattractor'' of this class of QDDM models. We denote with subindex $s$
magnitudes in the superattractor regime. As an example of models satisfying
$\zeta\to \zeta_s$ we note the case $\zeta\propto\sqrt{\rho_m}$, investigated
in Refs. \cite{Chi93,ChJMM,Coley94,Coley95}.

Let us characterize this asymptotic stage. From Eq. (\ref{gammapi}) the
superattractor solution yields a power--law evolution for the scale factor

\begin{equation} \label{asat}
a_{s}(t)=a_{0}\left( \frac{t}{t_{s}}\right) ^{\alpha } \,,
\end{equation}

\noindent where $\alpha=2/3\gamma_{\phi s}$, $a_0$ is the current scale factor
and $t_{s}$ is the age of a superattractor universe. 
The dynamics of the scalar field in the attractor regime is obtained from
Eqs. (\ref{gammaphi}) and (\ref{gammapi}):

\be
\n{Va}
V=\frac{3}{2}\Omega_{\phi a}(2-\gamma_\phi)\,H^2 \,,
\ee

\be
\n{pa}
\dot\phi^2=3\Omega_{\phi a}\gamma_{\phi}\,H^2 \,.
\ee

\no These equations together with (\ref{dHa}) close the problem of finding the
potential $V$ and the scalar field $\phi$ as functions of $t$. Expressing
the potential (\ref{Va}) and field (\ref{pa}) in terms of cosmological
time, that is, $V=\Omega_{\phi a}(3H^2+\dot H)$ and
$\dot\phi^2=-2\Omega_{\phi a}\dot H$, it is easy to find, for the
superattractor solution,

\be
\n{Vsa}
V_{s}(\phi)=\frac{2\Omega_{\phi a}}{3\left(\gamma_{\phi s}t_{s}\right)^2}
(2-\gamma_{\phi s}){\mbox e}^{-A\phi} \,,
\ee

\be
\n{psa}
\phi_{s}(t)=\phi_0+\sqrt{\frac{4\Omega_{\phi a}}{3\gamma_{\phi s}}}\,\ln
\frac{t}{t_{s}} \,,
\ee

\no where $A=(3\gamma_{\phi s}/\Omega_{\phi a})^{1/2}$ is the slope parameter.
Then we find $\alpha=2/(\Omega_{\phi a}A^2)$ and $\phi_0=(2/A)\ln\gamma_{\phi
s}$. We note that in the superattractor regime, the exponent $\alpha$ is
larger, by a factor of $1/\Omega_{\phi a}$, than the exponent $2/A^2$ of a
scalar field dominated era. This is a particular instance of a general
property of QDDM models in that dissipative effects assist quintessence driven
acceleration through the negative non-equilibrium pressure $\pi$. Using Eq.
(\ref{gammapi}) we can evaluate its ratio to quintessence pressure:

\begin{equation} \label{pipphi}
\frac{\pi}{p_{\phi}}=\frac{\gamma_\phi-\gamma_m}{\gamma_\phi-1}
\frac{\Omega_{ma}}{\Omega_{\phi a}} \,.
\end{equation}

\noindent Hence, in the superattractor regime, this ratio is also a constant
(provided that $\gamma_m$ is a constant), and when dark matter is cold it is
just the ratio of matter to quintessence energy density.

Let us sketch the evolution of the actual universe from a nearly
thermodynamical equilibrium early era (when $\pi/\rho_m\simeq 0$) into this
superattractor stage (when $\pi/\rho_m=\gamma_{\phi s}-\gamma_m$). Along the
radiation and matter dominated eras $|\pi/\rho_m|\ll 1$ and the inequality
$\gamma_m+\pi/\rho_m>\gamma_{\phi s}$ holds. Then dissipative processes become
more significant, the attractor condition (\ref{gammapi}) is approached,
$\gamma_\phi$ is driven towards $\gamma_{\phi s}$, and the expansion of the
scale factor accelerates. Clearly the timescale of this transition period
depends on the details of the dissipative processes occurring in dark matter,
encoded into the evolution of the dissipative magnitudes $\zeta(t)$ and
$\nu(t)$ -- some models exhibiting this convergence stage have been
investigated in Ref. \cite{cjp}. The superattractor stage finally settles when
$\zeta\simeq\zeta_s$. Assuming that the ratio of this transition period to the
age of the universe $t_0$ is small and this transition period started early
enough, we may approximate the recent evolution of $a(t)$ by the
superattractor solution (\ref{asat}); hence, $\gamma_{\phi 0} \simeq
\gamma_{\phi s}$ and $(1 + z)^{-1}=a/a_0\simeq (t/t_0)^\alpha$.

As Eq. (\ref{Vsa}) shows, the class of models converging to the superattractor
stage has $V(\phi)\sim V_s(\phi)$ for large $\phi$. In addition to this
exponential tail, no other constraint has to be imposed on the quintessence
potential for convergence to the superattractor era. For a wide range of
initial values of $\phi$ and $\dot{\phi}$ the quintessence field approaches a
common evolutionary path (\ref{psa}) for which $\gamma_\phi=\gamma_{\phi s}$;
i.e., the late behavior is insensitive to the initial conditions.

\section{Observational constraints}
 
\subsection{Luminosity distance of supernovae}

It has been found that  supernovae of type Ia (SNeIa) are nearly standard
candles. Properly corrected, the difference in their apparent magnitudes is
only related to differences in luminosity distance $d_L$ and consequently to
cosmological parameters \cite{Perlmutter98,Riess98}. Taking profit of this
property, several accelerated expanding cosmological models like $\Lambda$CDM
and QCDM have been fitted to recent observations of high redshift supernovae
($z\lsim 1$) \cite{Perlmutter98,Efstat,Perlmutter99,Wang99}. Though they have a
good fit in some regions of the parameter space corresponding to an
accelerated expansion, these models require fine tuning to account for the
observed ratio between dark energy and clustered matter.

On the other hand, QDDM models provide models that simultaneously provide an
accelerated expansion and solve the coincidence problem within the general
trend of the universe towards an attractor and with it towards the
superattractor. As in QCDM models, these scenarios depend on the quintessence
potential and initial conditions. In addition, they depend on the evolution of
the magnitudes characterizing the details of dissipative processes occurring
in dark matter. Let us examine the issue of reconstructing part of this
evolution through observations of distant SNeIa.

Ignoring gravitational lensing effects, the standard expression for the
luminosity distance to an object at redshift $z$ in a spatially homogeneous and
isotropic universe is \cite{Peebles93}

\be\label{dL}
d_L(z) = \frac{1+z}{H_0|\Omega_{k0}|^{1/2}~}{\cal S}\left(
H_0|\Omega_{k0}|^{1/2}
\int_0^z \frac{dz'}{H(z')}\right),
\ee

\noindent
with ${\cal S}(u)=(\sin u,u,\sinh u)$ for $k=(1,0,-1)$ respectively. Then for
a QDDM universe we have

$$
\frac{H(z)}{H_0}=
\left\{\Omega_{m0}\exp\left[{3\int^{z}_0\, \frac{dz'}
{1+z'}\left(\gamma_m+\frac{\pi}{\rho_m}
\right)}\right]\right.
$$
\begin{equation} \label{Hz}
\left.+\Omega_{\phi 0}\exp\left({3\int^{z}_0 \,\frac{dz'}
{1+z'}\,\gamma_{\phi}}\right)+\Omega_{k0} (1+z')^2\right\}^{1/2}
\end{equation}

In the case of a spatially flat universe, $d_L$ has a simpler expression in
terms of the effective baryotropic index

\begin{equation} \label{dLk0}
d_L(z) = \frac{1+z}{H_0}\int^z_0 dz'\exp\left[-{\frac{3}{2}\int^{z'}_0\,
\frac{dz''}{1+z''}\left(\gamma+\frac{\pi}{\rho}
\right)}\right]
\end{equation}

\noindent
where

\begin{equation} \label{gammak0}
\gamma+\frac{\pi}{\rho}=\gamma_m+\frac{\pi}{\rho_m}-
\left(\gamma_m+\frac{\pi}{\rho_m}-\gamma_\phi\right)
\Omega_{\phi 0}\exp\left({3\int^{z}_0 \,\frac{dz'}
{1+z'}\,\gamma_{\phi}}\right)
\end{equation}

\noindent We see from Eqs. (\ref{dLk0}) and (\ref{gammak0}) that $d_L(z)$
depends on the time evolution of dissipative processes through a double
integral in $\pi/\rho_m(z)$ and on the time evolution of the quintessence
field through a double or triple integral in $\gamma_\phi(z)$ (we assume that
$\gamma_m\simeq 1$). On one the hand, there is a degenerancy here as these two
functions cannot be reconstructed from knowledge of the single function
$d_L(z)$. Besides the time variation of these magnitudes is largely smoothed
out and, similarly to the analysis in Refs. \cite{IBS,Astier,Chevallier} for
QCDM models, we find that the luminosity distance is highly insensitive to
these variations. So we may safely replace these time-varying magnitudes by
their mean values in the interval $(0,z)$. Assuming that the universe has
already settled in the superattractor regime at the age of the farthest
observed supernova the degenerancy is eliminated and both functions become
constant. Then, the expression (\ref{dLk0}) simplifies drastically and we
obtain (cf. \cite{powerlaw})

\begin{equation} \label{dLsa}
d_L(z)=\frac{\left(1+z\right)\left[(1+z)^\beta-1\right]}{\beta H_0} \,,
\end{equation}

\noindent
where $\beta=a\ddot a/\dot a^2=1-1/\alpha$ is the acceleration parameter.
So $\beta$ increases with $\alpha$ and $1<\alpha<\infty$ corresponds to
$0<\beta<1$. On the superattractor we have

\be
\n{saa}
\gamma_{\phi s}={2\over 3\alpha}=\frac{2}{3}(1-\beta) \,.
\ee

We have used the sample of $38$ high redshift ($0.18 \le z \le 0.83$)
supernovae of Ref. \cite{Perlmutter98}, supplemented with $16$ low redshift
($z < 0.1$) supernovae from the Cal\'an/Tololo Supernova Survey
\cite{Hamuy}. This is described as the ``primary fit'' or fit C in Ref.
\cite{Perlmutter98}, where, for each supernova, its redshift $z_i$, the
corrected magnitude $m_i$ and its dispersion $\sigma_i$ were computed.

The predicted magnitude for an object at redshift $z$ with luminosity distance
$d_L$ is

\begin{equation}
m(z) =  {\cal M} + 5\log{\cal D}_L(z),
\end{equation}

\no where ${\cal M}$ is related to the absolute magnitude $M$ by

\begin{equation} \label{}
{\cal M} = M- 5 \log\left(\frac{H_0[\mbox{\rm km/s Mpc}]}
{c[\mbox{\rm km/s}]}\right) + 25
\end{equation}

\noindent and ${\cal D}_L$ is the  luminosity distance in units of
the Hubble radius.

We have determined the optimum fit of the superattractor model by minimizing a
$\chi^2$ function:

\begin{equation} \label{chi2}
\chi^2=\sum^N_{i=1}\frac{\left[m_i-m(z_i;\beta,{\cal M})\right]^2}
{\sigma_i^2} \,,
\end{equation}

\noindent where $N=54$ for this data set. The most likely parameters are
$(\beta,{\cal M})=(0.395,23.96)$, yielding $\chi^2_{min}/N_{DF}=1.12$
($N_{DF}=52$), and a goodness--of--fit $P(\chi^2\ge \chi^2_{min})=0.253$.
These numbers show that the fit of superattractor QDDM cosmology to this data
set is as good as the fit of the $\Lambda$CDM model (see Fig. 1). We note that
it occurs even though the superattractor model has basically only the
acceleration parameter to fit large redshift supernovae (${\cal M}$ being
largely determined by low redshift supernovae). It also means that the density
of clustered matter is not constrained by measurements of SNeIa and it has to
be determined through independent observations.

We estimate the probability density distribution of the parameters by
evaluation of the normalized likelihood \cite{Lupton93}

\begin{equation} \label{pbetaM}
p(\beta,{\cal M})=\frac{\exp\left(-\chi^2/2\right)}
{\int d\beta \int d{\cal M}\exp\left(-\chi^2/2\right)} \,.
\end{equation}

\noindent Then we obtain the probability density distribution for $\beta$
marginalizing $p(\beta,{\cal M})$ over ${\cal M}$. This probability density
distribution $p(\beta)$ is plotted in Fig. 2 and it yields $\beta=0.398\pm
0.104$ $(1\sigma)$. Hence we can state that $0.085<\beta<0.711$ with a
confidence level of $0.997$, so that an accelerated superattractor QDDM
universe is strongly supported by this data set, in agreement with a similar
analysis of $\Lambda$CDM and QCDM models
\cite{Perlmutter98,Riess98,Garnavich98,Perlmutter99}.

\subsection{Age of the Universe}

For a QDDM model the age of the universe has the integral representation

\begin{equation} \label{t0}
t_0=\int^\infty_0 \frac{dz}{\left(1+z\right)H(z)} \,,
\end{equation}

\noindent where $H(z)$ is given by Eq. (\ref{Hz}). Along the transition period of
the universe from an early nearly equilibrium stage towards the superattractor
stage the inequality $\gamma_m+\pi/\rho_m>\gamma_{\phi s}$ holds. As a
consequence $H(z)>H_{s}(z)=H_0(1+z)^{1-\beta}$ (whenever $\Omega_k\ge 0$)
and $t_0<t_{s}=\alpha/H_0$. The characteristics of this transition period, and
hence the value of the difference $t_{s}-t_0$, are model dependent and will
not be dealt with here.

Using the fit to SNe Ia of the previous subsection and Eq. (\ref{saa}) we
obtain $\alpha=1.711\pm0.288$. Combined with $H_0=65\pm 5$ km/s Mpc
\cite{Turner00}, it yields for the age of the superattractor universe
$t_{s}=25.9\pm 4.8$ Gyr. This shows that the QDDM cosmology does not suffer of
any age discrepancy and can accommodate comfortably the $1\sigma$ interval
$9$--$16$ Gyr for the age estimate of globular clusters \cite{Primack00}.

\subsection{Parameters of the superattractor era}

Using the probability distribution for $\beta$ and Eq. (\ref{saa}) we obtain
$\gamma_{\phi s}=0.401\pm 0.069$, in agreement with previous results for QCDM
models in the limit $\Omega_m\to 0$. Then, using Eq. (\ref{gammapi}) and
assuming that dark matter is cold ($\gamma_m=1$), we find
$\pi/\rho_m=-0.599\pm 0.069$. This figure implies that substantial dissipative
processes are taking place in dark matter. This fact is also shown by the
large value of the effective baryotropic index. In effect, combining the
estimate $\Omega_{m0}=0.35\pm 0.07$ from cluster baryons \cite{Turner00} with
the a priori constraint $\Omega=1$ and inserting into Eq. (\ref{gammaOmega})
we get $\gamma_s=0.61\pm 0.09$. For a perfect fluid this value would
correspond to a power-law exponent $2/3\gamma_s\simeq 1.1$, quite lower than
$\alpha$. Using Eq. (\ref{vis}), we find that the linear relationship
$\kappa=\kappa_1+\kappa_2\nu^{-1}$ holds, where $\kappa_1=0.21\pm 0.05$ and
$\kappa_2=-0.25\pm 0.07$. The requirement of asymptotic stability of the
superattractor solution imposes that $\nu>1.20\pm 0.21$.

Using Eqs. (\ref{Va}) and (\ref{pa}), we find the quintessence kinetic energy
density parameter $\Omega_{K0}=0.13\pm 0.03$ and $\Omega_{V0}=0.52\pm 0.06$
for the potential energy density parameter. These figures show that the scalar
field is moving down the potential outside the slow--roll regime.
Similarly we find for the slope of the exponential potential $A=1.36\pm 0.14$
and for the current value of the scalar field $\phi_0=-1.3\pm 0.3$. This
implies a mass parameter of the Planck scale.

\section{Density perturbations}

Besides the model degenerancy of luminosity distance determinations, even
within the superattractor cosmology there are some thermodynamical parameters
like $\nu$ and the speed of sound $c_{\rm s}$ that are not fixed. For this
reason we will investigate the evolution of density fluctuations in the
perturbative long--wavelength regime during the superattractor era. It is
possible that weak lensing techniques could yield more information about these
parameters \cite{Kaiser}.

Scalar perturbations are covariantly and gauge--invariantly characterized by
the spatial gradients of scalars. Density inhomogeneities are described by
the comoving fractional density gradient \cite{eb}

\begin{equation}
\delta _{i}={\frac{a\D_{i}\rho }{\rho }} \,,  \label{e'}
\end{equation}

\noindent where $\D_{i}$ stands for the covariant spatial derivative $\D
_{j}A_{i\cdots }=h_{j}{}^{k}h_{i}{}^{l}\cdots \nabla _{k}A_{l\cdots }$. The
scalar part $\delta \equiv a\D^{i}\delta _{i}=(a\D)^{2}\rho /\rho $
corresponds to the usual gauge-invariant density perturbation scalar $\ep_{
\rm m}$ \cite{b,bde}, which encodes the total scalar
contribution to density inhomogeneities. Also the comoving expansion
gradient, the normalized pressure gradient, and normalized entropy gradient
are defined by \cite{eb,mt}

\begin{equation}
\theta _{i}=a\D_{i}\theta \,,\quad
p_{i}={\frac{a\D_{i}p}{\rho }} \,,\quad
e_{i}={\frac{a n T\D_{i}s}{\rho }}\, ,  \label{e''}
\end{equation}

\noindent  $n$ being the particle number density, $T$ the temperature, and $s$
the specific entropy per particle. The evolution equation for scalar density
perturbations reads \cite{mt}

\[
\ddot{\delta}+H\left( 8-6\gamma +3\cs\right) \dot{\delta}-{\ts{3\over2}}
H^{2}\left\{ 1+5\left( \gamma -1\right) ^{2}-6\cs+\right.
\]
\begin{equation}
\left.\left[( 1-3\left( \gamma
-1\right) ^{2}+2\cs\right] k\right\} \delta
-\cs\D ^{2}\delta ={\sf S}[e]+{\sf S}[\pi]+{\sf S}[q]+{\sf S}
[\sigma ]\,,  \label{l}
\end{equation}

\noindent where

\begin{equation}  \label{cs}
\cs=\left( {\frac{\p p}{\p\rho }}\right) _{\!s} \,,\quad
r={\frac{1}{nT}}\left( \frac{\p p}{\p s}\right) _{\!\rho },
\end{equation}

\noindent are, respectively, the adiabatic speed of sound and a
non-baryotropic index. The sources in the right--hand side of Eq. (\ref{l})
arising, respectively, from entropy perturbations, bulk viscous stress, energy
flux, and shear viscous stress are given in \cite{mt}. Since in our case there
are no shear viscous stress ($\sigma _{ij}=0$) and ${\sf S}[q]$ vanishes by
choosing the energy frame ($q_{i}=0$), we reproduce here only the expressions
for $ {\sf S}[e]$ and ${\sf S}[\pi]$:

\begin{eqnarray}
{\sf S}[e] &=&r\left( 3k H^{2}+\D^{2}\right) e\,,  \label{l1} \\
{\sf S}[\pi] &=&-\left( 3k H^{2}+\D^{2}\right) {\cal B}\,,  \label{l2}
\end{eqnarray}

\noindent where the scalar entropy perturbation

\begin{equation}
e=a\D^{i}e_{i}={\frac{a^{2}nT}{\rho }}\D^{2}s
\end{equation}

\noindent and the dimensionless perturbation scalar

\begin{equation}
{\cal B}={\frac{a^{2}\D^{2}\pi }{\rho }}\,,~  \label{l5}
\end{equation}

\noindent related to the inhomogeneous part of the bulk viscous stress, were
defined.

Also, the entropy perturbation equation in the energy frame is

\begin{equation}
\dot{e}+3H\left( \cs-\gamma +1+r\right) e=-3H {\cal B}\ .  \label{entr}
\end{equation}

The coupled system that governs scalar dissipative perturbations in the
general case is given by the density perturbation equation (\ref{l}), the
entropy perturbation equation (\ref{entr}), the equation for the scalar bulk
viscosity (\ref{dpi2}), and the equation for temperature perturbations.

When only bulk viscous stress dissipation is present, the coupled system can
be reduced to a pair of coupled equations in $\delta $ (third order in time)
and $e$ (second order in time). For a flat background, the equations are
\cite{mt}:

\[
\tau  \stackrel{\ldots }{\delta }+\left[ 1+3\left( 2-\gamma +\cs\right)
\tau  H\right] \ddot{\delta}+H\left\{ 8-6\gamma +3\cs+3\tau  \left( \cs \right)
^{\rd}\right.
\]
\[
\left. -{\ts{1\over2}}\left[ -14+75\gamma -48\gamma
^{2}+(21\gamma -30)\cs \right] \tau  H\right\} \dot{\delta}
\]
\[
-{\ts{3\over2}}H^{2}\left\{ 6-10\gamma +5\gamma ^{2}-6\cs-4\tau  \left( \cs
\right) ^{\rd }\right.
\]
\[
\left. -2\left[ -6+18\gamma -15\gamma
^{2}+5\gamma ^{3}+\left( 6-28\gamma +10\gamma ^{2}\right)  \cs\right] \tau
H\right\} \delta
\]
\[
={\frac{a^{2}\zeta }{\rho  \gamma }}\D^{2}\left(
\D^{2}\dot{\delta}\right) + {\frac{3a^{2}(\gamma -1)H}{\rho  \gamma
}}\D^{2}\left( \D^{2}\delta \right) +\tau  \cs\D^{2}\dot{\delta}+\tau
r\D^{2}\dot{e}
\]
\[
+\left[ \left( 1-3\gamma  \tau  H\right) \cs+\tau
\left( \cs \right) ^{\rd}+3\left( {\frac{\p\zeta }{\p\rho }}\right)
_{\!s}\right] \D ^{2}\delta
\]
\begin{equation}
+\left[ \left( 1-3\gamma
\tau  H\right) r+\tau  \dot{r}+3\left( {\frac{ \p\zeta }{\p s}}\right)
_{\!\rho }\right] \D^{2}e\,,
\label{d2} \end{equation}

\noindent and

\[
\tau  \ddot{e}+\left[ 1-{\ts{3\over2}}\left( -2+3\gamma -2\cs-2r\right)
\tau  H\right] \dot{e}
\]
\[
-3H\left[ \gamma -1-\cs-r+3\gamma  \left( \gamma -\cs\right) \tau  H+\tau
 \left( \cs
+r\right) ^{\rd}\right.
\]

\begin{equation}
\left. -{\frac{\rho }{\gamma }}\left( {\frac{\p\zeta }{\p s}}\right)
_{\!\rho }\right] e=-{\frac{\zeta }{\gamma }\ }\dot{\delta}+{\frac{3H}{
\gamma }}\left[ \left(\gamma-1\right)\zeta +\rho
\left( {\frac{\p\zeta }{\p\rho }}\right)
_{\!s}\right] \delta \,.  \label{d3}
\end{equation}

We shall consider here the evolution of the density and entropy perturbations
in the superattractor stage with the conditions $r=0$ and $\partial \nu
/\partial s=0$. Together with Eq. (\ref{vis}) they imply

\begin{eqnarray}
\left( \frac{\partial \zeta }{\partial s}\right) _{\rho } &=&0\ , \\
\left( \frac{\partial \zeta }{\partial \rho }\right) _{s} &=&\left( \frac{
\partial \zeta }{\partial \rho _{m}}\right) _{s}=\frac{ \kappa
_{1}+\kappa _{2}/\nu }{6\Omega _{m}H}
\end{eqnarray}

\noindent
In this case, Eq. (\ref{d2}) decouples to give
\begin{eqnarray}
\stackrel{\ldots }{\delta }+\frac{c_{1}}{t}\ddot{\delta}+\frac{c_{2}}{t^{2}}
\dot{\delta}+\frac{c_{3}}{t^{3}}\delta &=&c_{4}\ t^{2\alpha }\ \D^{4}\dot{
\delta}+c_{5}\ t^{2\alpha }\ \D^{4}\delta +\cs \D^{2}\dot{\delta}
\nonumber \\
&&+\left( \frac{c_{6}}{t}+c_{7}\right) \D^{2}\delta \ ,  \label{delta}
\end{eqnarray}

\noindent and Eq. (\ref{d3}) becomes
\begin{equation}
\ddot{e}+\frac{c_{8}}{t}\ \dot{e}+\frac{c_{9}}{t^{2}}\ e=\frac{c_{10}}{t^{2}}
\ \dot{\delta}+\frac{c_{11}}{t^{2}}\ \delta  \label{entrop}
\end{equation}

\noindent where the constant coefficients $c_{1}\ldots c_{11}$ depend upon the
parameters of the model: $\nu $, $\kappa _{1}$, $\kappa _{2}$, $\Omega
_{m},\alpha $, $\gamma $, $\cs$, and the present value of the scale factor
$a_{0}$. For our purposes, only the explicit expression for
$c_{1}$, $c_8$, and $c_9$ are
relevant, being

$$
c_{1}=\alpha \nu +3\alpha \ \left( 2-\gamma -\cs \right) \,,
$$
$$
c_8=\alpha\left[\nu-\frac{3}{2}\left(3\gamma-2-2\cs\right)\right] \,,
$$
\begin{equation}\label{c1}
c_9=-3\alpha^2\left[\nu\left(\gamma-1-\cs\right)+
3\gamma\left(\gamma-\cs\right)\right] \,.
\end{equation}

\noindent We deal with the system (\ref{delta}),(\ref{entrop}) by performing
separation of variables  in the form $\delta =\delta _{x}\ \delta _{t}$ and $
e=e_{x}\ e_{t} $, where $\delta _{x}$ and $e_{x}$ depend upon the spatial
variables while $\delta _{t}$ and $e_{t}$ are functions of the coordinate time
$t.$ Then, Eq. (\ref{delta}) can be recasted as

\begin{eqnarray}
\frac{\stackrel{\ldots }{\delta }_{t}}{\delta _{t}}+\frac{c_{1}}{t}\frac{
\ddot{\delta}_{t}}{\delta _{t}}+\frac{c_{2}}{t^{2}}\frac{\dot{\delta}_{t}}{
\delta _{t}}+\frac{c_{3}}{t^{3}} &=&t^{2\alpha }\left( c_{4}\frac{\dot{
\delta }_{t}}{\delta _{t}}+c_{5}\right) \frac{\D^{4}\delta _{x}}{\delta _{x}}
\nonumber\\
&&+\left( \cs \frac{\dot{\delta}_{t}}{\delta _{t}}+\frac{c_{6}}{t}
+c_{7}\right) \frac{\D^{2}\delta _{x}}{\delta _{x}}\ ,
\end{eqnarray}

\noindent which can only hold if

\begin{equation}
\left( \D^{2}-\mu \right)  \delta _{x}=0\ ,  \label{proca}
\end{equation}

\noindent  $\mu $ being an arbitrary constant. Also, Eq. (\ref{entrop})
leads to

\begin{equation} \label{deltae}
\frac{t^{2}\left( \ddot{e}_{t}/e_{t}\right) +c_{8}\ t\ \left( \dot{e}
_{t}/e_{t}\right) +c_{9}}{c_{10}\left( \dot{\delta}_{t}/e_{t}\right)
+c_{11}\left( \delta _{t}/e_{t}\right) }=\frac{\delta _{x}}{e_{x}}\ ,
\end{equation}

\noindent which requires $e_{x}=A\ \delta _{x}\ $, with $A$ an arbitrary
constant which can be absorbed into the temporal functions, resulting in the
same spatial distribution of the entropy and density perturbations. Under
the condition (\ref{proca}) the evolution equation for $\delta _{t}$ becomes

\[
\stackrel{\ldots }{\delta }_{t}+\frac{c_{1}}{t}\ddot{\delta}_{t}+\left(
\frac{c_{2}}{t^{2}}-\mu ^{2}\ t^{2\alpha }\ c_{4}-\mu \cs \right) \dot{
\delta }_{t}
\]
\begin{equation}  \label{deltat}
+\left[ \frac{c_{3}}{t^{3}}-\mu ^{2}\ t^{2\alpha }\ c_{5}-\mu \left( \frac{
c_{6}}{t}+c_{7}\right) \right] \delta _{t} =0\ .
\end{equation}

As we are interested in the asymptotic behavior for the superattractor regime,
it suffices to consider the dominant terms in (\ref{deltat}),  $\alpha $ being
a positive number, we have

\begin{equation}
\stackrel{\ldots }{\delta }_{t}+\frac{c_{1}}{t}\ddot{\delta}_{t}-\mu ^{2}\
t^{2\alpha }\ c_{4}\ \dot{\delta}_{t}-\mu ^{2}\ t^{2\alpha }\ c_{5}\ \delta
_{t}\cong 0\ .  \label{asymptotic}
\end{equation}

Equations (\ref{proca}) and (\ref{asymptotic}) give the form of the
asymptotic density perturbations in the model. The parameter $\mu $ appearing
in (\ref{proca}) depends, in principle, on the boundary conditions of the
problem being the quantity $\mu ^{-1/2}$, a characteristic coordinate length
related to the range of the exponentially decaying modes of the spatial part
$\delta _{x}.$ In the special case $\mu =0$, Eq. (\ref{proca}) has the
Laplace form, describing long--range density perturbation modes. Here, we
are going to study the asymptotic evolution of the long--range modes by
performing a series expansion of $\delta _{t}$ in powers of $\mu ^{2}.$ Up
to first order we have $\delta _{t}\cong \delta _{t}^{(0)}+\mu ^{2}\ \delta
_{t}^{(1)},$ then, replacing this expression in Eq. (\ref{asymptotic}) and retaining terms up to first order in $\mu ^{2}$ we obtain

\begin{equation}\label{deltat1}
\left[ \stackrel{\ldots }{\delta }_{t}^{(1)}+\frac{c_{1}}{t}\ddot{\delta}
_{t}^{(1)}-t^{2\alpha }\ c_{4}\ \dot{\delta}_{t}^{(0)}-t^{2\alpha }\ c_{5}\
\delta _{t}^{(0)}\right] \ \mu ^{2}+\stackrel{\ldots }{\delta }_{t}^{(0)}+
\frac{c_{1}}{t}\ddot{\delta}_{t}^{(0)}=0\ .
\end{equation}

\noindent The zeroth--order solution gives

\begin{equation}
{\delta}_{t}^{(0)}(t)=\frac{A_{1}}{(1-c_{1})(2-c_{1})}\
t^{2-c_{1}}+A_{2}\ t+A_{3}\ ,  \label{delta0}
\end{equation}

\noindent for $c_{1}\notin \{1,2\}$ and $A_{i}$ , $i=1,2,3$ integration
constants. Then, $\delta _{t}^{(1)}$ satisfies the inhomogeneous equation

\begin{equation} \label{deltat2}
\stackrel{\ldots }{\delta }_{t}^{(1)}+\frac{c_{1}}{t}\ddot{\delta}
_{t}^{(1)}-t^{2\alpha }\ c_{4}\ \dot{\delta}_{t}^{(0)}-t^{2\alpha }\ c_{5}\
\delta _{t}^{(0)}=0\ ,
\end{equation}

\noindent whose general solution has the form

\begin{eqnarray}
\delta _{t}^{(1)} &=&\frac{B_{1}}{(1-c_{1})(2-c_{1})}\ t^{2-c_{1}}+B_{2}\
t+B_{3}+a_{1}t^{2\alpha +4-c_{1}}+ \nonumber\\
&&+a_{2}\ t^{2\alpha +3}+a_{3}\ t^{2\alpha +5-c_{1}}+a_{4}\ t^{2\alpha +4}\ ,
\end{eqnarray}

\noindent where the coefficients $B_{i}$, $i=1,2,3$ are integration constants
depending upon the initial conditions, and $a_{i}$, $i=1,\ldots ,4$ depend on
$\alpha $, $c_{1}$, and the $B_{i}$. Then, the asymptotic evolution of the
long range modes up to first order in $\mu ^{2}$ results in a combination of
powers of time. When $c_{1}<2$ all the exponents are positive, while for
$c_{1}>2$ there exists a decaying mode. With the parameters of the model,
because of Eq. (\ref{c1}) this last situation is equivalent to $\cs <\nu/
3+0.999$ which is always satisfied because of the lower bound $\nu>1.2 $ (in
our units $\cs <1$). As a consequence, a decaying mode exists in this model
and the dominant exponent is $2\alpha+4$. Hence the density power spectrum
redshifts as $P(k,z)\propto (1+z)^{-(4+8/\alpha)}$.

On the other hand, Eq. (\ref{d3}) becomes, in the leading regime,

\begin{equation} \label{de}
\ddot e_t+\frac{c_8}{t}\dot e_t+\frac{c_9}{t^2}e_t=
c_{12}t^{2\alpha+1}\left[1+\frac{3}{\nu}\left(\gamma-1+
\frac{1}{2\Omega_{ma}}\right)t\right] \,,
\end{equation}

\noindent
whose solution is

\begin{equation} \label{et}
e_t(t)=B_4 t^{\lambda_1}+B_5 t^{\lambda_2}+c_{13}
t^{2\alpha+3}\left[1+\frac{3}{\nu}\left(\gamma-1+
\frac{1}{2\Omega_{ma}}\right)t\right] \,,
\end{equation}

\noindent where $\lambda_{1,2}$ are the roots of the equation
$\lambda^2+(c_8-1)\lambda+c_9=0$, $B_4, B_5$ are arbitrary integration
constants, and $c_{12},c_{13}$ are functions of the parameters and the
previously defined integration constants. In this way the entropy
perturbations may yield additional information about the parameters $\nu$,
$\cs$, and $\Omega_{ma}$.

\section{Discussion}

We have shown that within the class of accelerated coincidence--solving
attractor solutions of QDDM models, a distinguished  attractor solution exists
with constant quintessence baryotropic index. This superattractor cosmology is
quite attractive because of its simplicity: the scale factor expansion follows
a power--law, and all its parameters can be evaluated in a model independent
way. Notwithstanding its simplicity, its fit to SNeIa observations is as good
as the $\Lambda$CDM model, and it does not suffer from any age conflict
either.

Our results are quite valuable to give general statements about a large class
of QDDM models that are driven towards this superattractor but they are
limited to the late--time regime. Undoubtedly a large variety of models arise
both from different quintessence potentials as well as from different
dissipative processes in dark matter. The only requirement is that the
potential has the exponential tail (\ref{Vsa}) or, equivalently, that the
viscosity coefficient has the asymptotic behavior (\ref{vis}). This diversity
of possibilities makes the transition period of the universe from its nearly
thermodynamical equilibrium early stage towards the superattractor regime
model dependent and requires more detailed investigation. In particular it
would be of importance to evaluate the effects of dissipative processes on the
CMB angular power spectrum.

Standard CDM models have been quite successful in describing the evolution of
cosmic structure. Hence a sufficiently long matter--dominated era must have
taken place during which the observed structure grew from the density
fluctuations measured by CMB anisotropy experiments. As a consequence the
transition to the currently observed accelerated regime should have been quite
recent in the evolution of the universe. For QDDM models this seems to imply
that dissipative effects in dark matter were small until density
inhomogeneities became large. If so, it is suggestive to think that the size
of dissipative processes, as measured by the ratio $\pi/\rho_m$, grows with
dark matter density and became large precisely because of the development of
inhomogeneities. The observed smoothness of halo structure and the correlation
between high energy cosmic ray production and dark matter clustering  might
provide support for this relationship.

Even though the supernovae search  is extended to $z>1$ this does not enable a
precise determination of the time variation in the quintessence baryotropic
index $\gamma_\phi$ and the ratio $\pi/\rho_m$ because the luminosity distance
depends on these magnitudes through a multiple-integral relation that  smears
out detailed information about their variability. Besides their independent
variation cannot be disentangled from the determination of a single function.
Hence further independent cosmological probes are required to investigate this
issue. As a first step in this direction we have calculated the dominant
large--time long--wavelength behavior of density and entropy fluctuations in
the superattractor regime.

The combined distribution of dark mass and dark energy can be investigated via
its gravitational effects inducing correlated shear in the images of distant
galaxies. This weak lensing effect could, in principle, be used to probe the
large--scale structure and thereby yield additional information about the
thermodynamical and cosmological parameters not constrained by the
luminosity--redshift relationship.

\section*{Acknowledgments}

This work has been supported by the University of Buenos Aires under
project TX-93.



\begin{figure}[thbp]
\plot{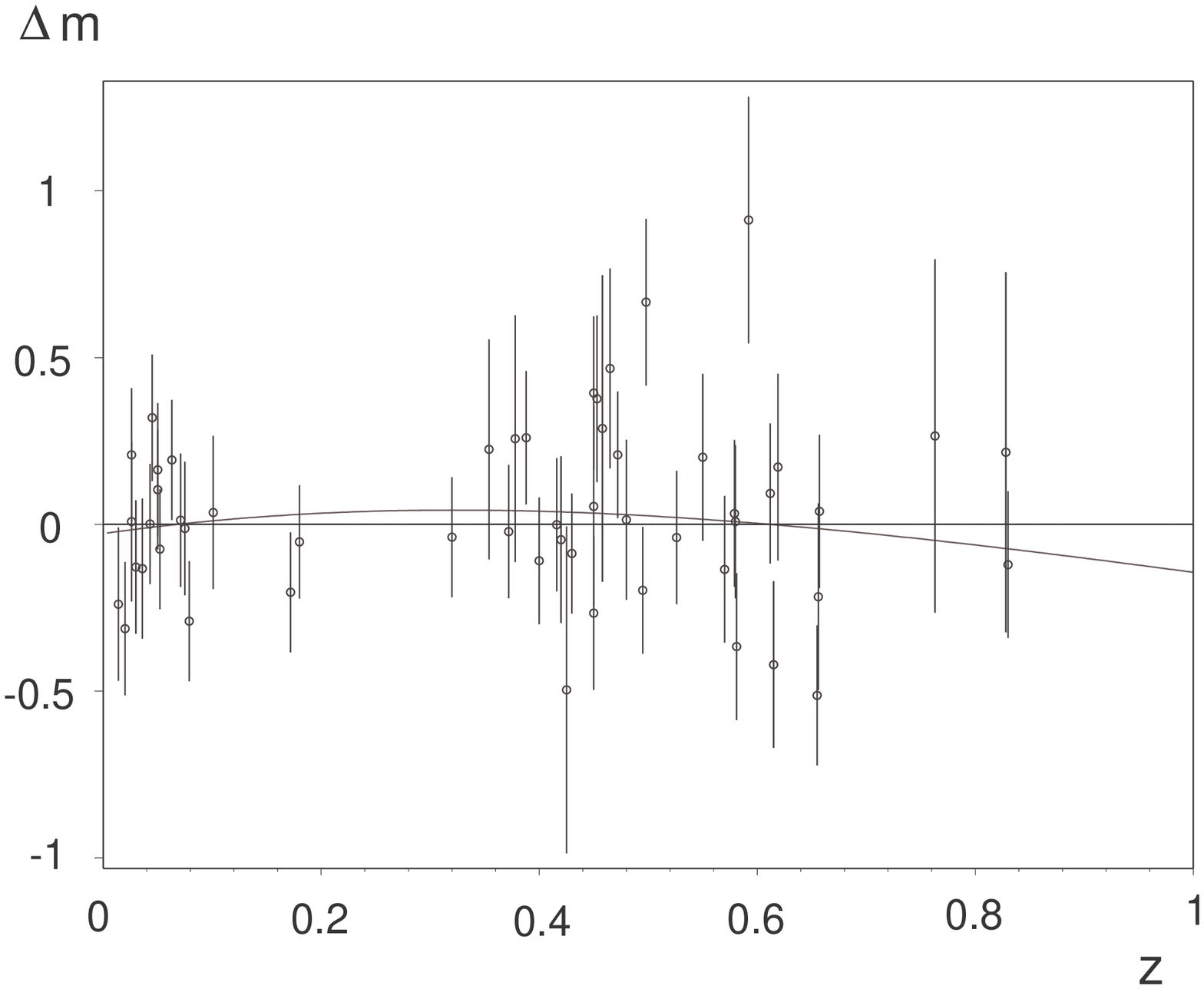}
\caption[caption]{
The magnitude residuals $\Delta m$ of the SNeIa (fit C data set of Ref.
\cite{Perlmutter98}) from the best fit superattractor QDDM model with
$(\beta,{\cal M})=(0.395,23.96)$. For comparison is shown the magnitude
residual for the best fit $\Lambda$CDM model, corresponding to
$(\Omega_{m0},\Omega_\Lambda,{\cal M})=(0.54,1.09,23.93)$ \cite{Efstat}.
}
\label{fig1}
\end{figure}


\begin{figure}[thbp]
\plot{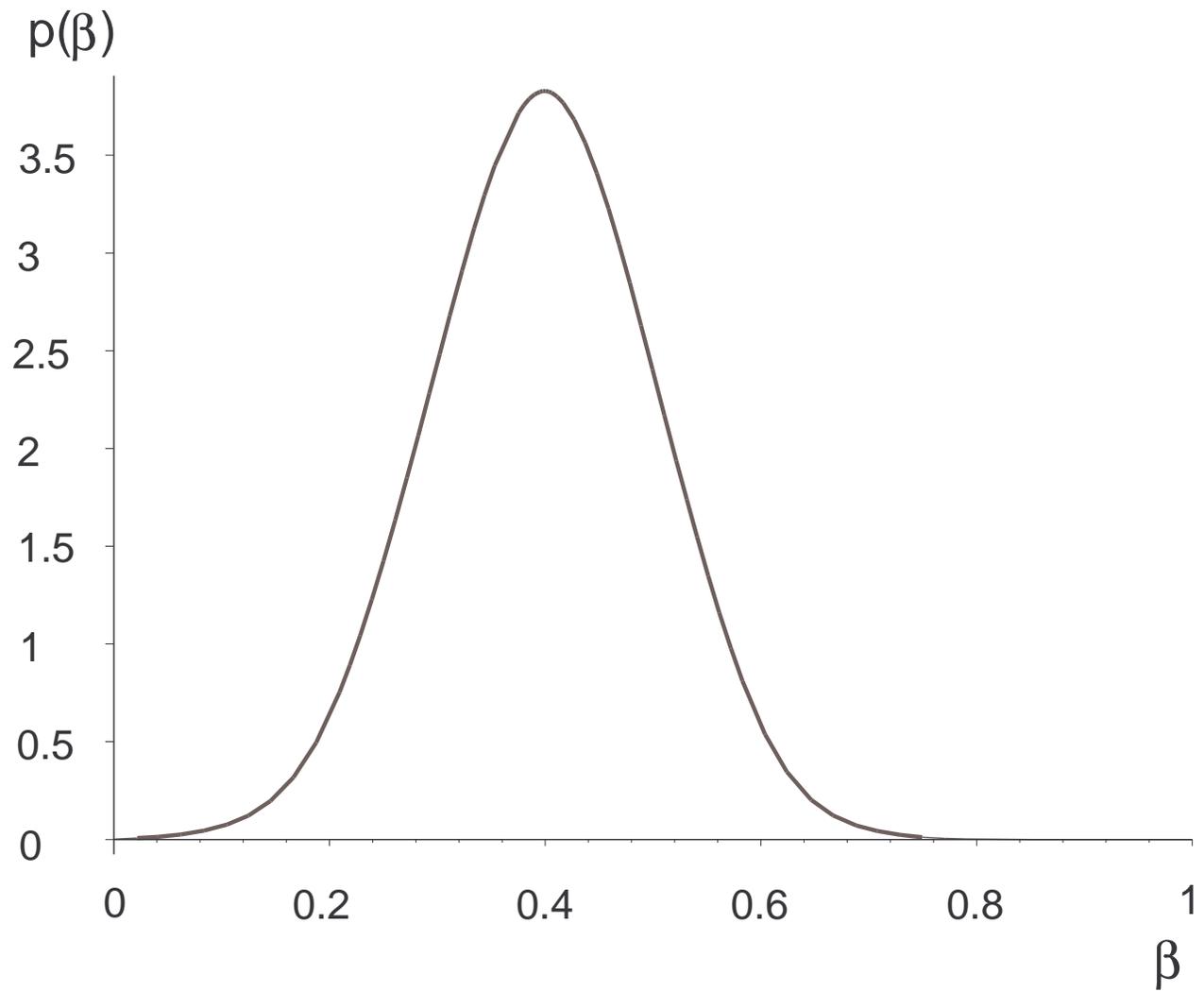}
\caption[caption]{
The estimated probability density distribution (normalized likelihood)
for the acceleration parameter $\beta$ of the superattractor model.
}
\label{fig2}
\end{figure}


\end{document}